\documentclass{PoS}

\usepackage{latexsym}
\usepackage{subfigure}
\usepackage{multicol}
\usepackage{multirow}
\usepackage{mathrsfs}
\usepackage{verbatim}
\usepackage{amsmath,amsthm, amssymb}

\newcommand{\stat}{\textrm{stat}\,}
\newcommand{\kin}{\textrm{kin}\,}
\newcommand{\spin}{\textrm{spin}\,}

\newcommand{\HQET}{\textrm{HQET}\,}
\newcommand{\QCD}{\textrm{QCD}\,}

\title{Fixing the parameters of Lattice HQET including $1/m_B$ terms}

\ShortTitle{Fixing the parameters of Lattice HQET including $1/m_B$ terms}

\author{
	\speaker{Piotr Korcyl} 	
\hfill{\footnotesize{\it DESY 13-103}}\\
        NIC, DESY, Platanenallee 6, 15738 Zeuthen, Germany\\
        E-mail: \email{piotr.korcyl@desy.de}}

\author{(for the ALPHA collaboration)}

\abstract{The study of the CKM matrix elements with increasing precision requires 
  a reliable evaluation of hadronic matrix elements of axial and vector 
  currents which can be done with Lattice QCD. The tiniest entry, $|V_{ub}|$, 
  can be estimated independently from $B \rightarrow \tau \nu$ and $B \rightarrow \pi l \nu$ decays. 
  The ALPHA collaboration has undertaken the 
  effort to evaluate non-perturbatively the decay constant $f_B$ and the $f^+(q)$ form 
  factor for $q^2$ close to $q_{\textrm{max}}^2$ entering these determinations. Since for the \textbf{b} quark 
  $m_b \gg a^{-1}$ for available lattice sizes,
  an effective description of the \textbf{b} quark is necessary. HQET provides an example of the latter. 
  As any effective theory, HQET is predictive only when a set of parameters have 
  been determined through a process called matching. 
  The non-perturbative matching procedure applied by the 
  ALPHA collaboration consists of 19 matching conditions 
  needed to fix all the relevant parameters at order $1/m_B$ of the HQET action and the axial and vector currents. 
  We present a study of one-loop corrections to two representative matching conditions.
  Our results enable us to quantify the quality of the observables used in the matching procedure. 
}

\FullConference{14th International Conference on B-Physics at Hadron Machines,\\
		April 8-12, 2013\\
		Bologna, Italy }

\begin{document}

Heavy Quark Effective Theory \cite{eichten_hill} in its basic formulation provides an effective description of QCD 
with $N_f-1$ light quarks and a single heavy quark whose mass is much larger 
than the QCD energy scale $\Lambda_{\QCD}$. The heavy quark is treated non-relativistically and 
processes are described in its reference frame. In order to avoid the ambiguities of the perturbative 
expansion of HQET in the strong coupling $g$ \cite{grozin} one may employ non-perturbative techniques such as lattice QCD 
and consequently lattice HQET \cite{leshouches}. As any effective theory HQET containes several low energy constants
which need to be determined in order to match it to QCD and grant it a predictive power. This step called 'matching'
should also be performed non-perturbatively \cite{sachrajda}. The ALPHA collaboration has set up a 
non-perturbative matching strategy to determine the needed HQET parameters at order $1/m_b$ \cite{heitger_sommer,1,2}. 
It relies on a set of carefully chosen observables which are precisely 
computable in lattice QCD as well as in lattice HQET. We describe the results of a one-loop computation which tests the quality
of some of these observables. In order to estimate the $1/m_b^2$ contributions we define a quantity $R$ which
measures the ratio of the one-loop corrections to their tree-level value of $1/m_b$ terms. The paper is organized 
as follows: in section \ref{sec1} we introduce the lattice HQET Lagrangian 
and the currents as well as higher dimensional operators needed to account for $1/m_b$ corrections, then in section \ref{sec2} 
we briefly describe the framework in which the matching observables are constructed and finally in section \ref{sec3} 
we discuss the one-loop results and conclude.

\section{HQET at next-to-leading order in $m_b$}
\label{sec1} 

The formulation of lattice HQET was thoroughly discussed in \cite{leshouches} and therefore we only quote the 
relevant formulae. 
The Lagrangian is a sum of the static part and two $1/m_b$ corrections
\begin{equation}
\mathcal{L}_{\HQET} = \mathcal{L}_{\stat} - \Big( {\color{black} \omega_{\kin}} \mathcal{L}_{\kin} + {\color{black} \omega_{\spin}} \mathcal{L}_{\spin} \Big) + \mathcal{O}(1/m_b^2),
\end{equation}
It is part of the definition of HQET that the kinetic and chromomagnetic operators enter 
only as insertions in the static vacuum expectation values, namely for some operator $\mathcal{O}$ we have
\begin{equation}
\langle \mathcal{O} \rangle_{\HQET} = \langle \mathcal{O} \rangle_{\textrm{stat}}
+ {\color{black}\omega_{\textrm{kin}}} \sum_{x} \langle \mathcal{O} \mathcal{L}_{\textrm{kin}}(x)
\rangle_{\textrm{stat}}
+ {\color{black}\omega_{\textrm{spin}}} \sum_{x} \langle \mathcal{O} \mathcal{L}_{\textrm{spin}}(x)
\rangle_{\textrm{stat}} 
\end{equation}
The HQET operators themselves are also expanded in $1/m_b$. 
For the axial current we have
\begin{align}
\big(A_0\big)_R &= {\color{black}Z_{A_0}^{\HQET}} \big\{ \bar{\psi}_l \gamma_0 \gamma_5 \psi_h
+ {\color{black} c_{A_{0,1}}} \bar{\psi}_l \frac{1}{2} \gamma_5 \gamma_i ( \nabla_i^S - \overleftarrow{\nabla}_i^S) \psi_h
+ {\color{black} c_{A_{0,2}}} \bar{\psi}_l \frac{1}{2} \gamma_5 \gamma_i ( \nabla_i^S + \overleftarrow{\nabla}_i^S) \psi_h
 \big\}, \nonumber  \\
%
\big(A_k\big)_R &= {\color{black}Z_{A_k}^{\HQET}} \big\{ \bar{\psi}_l \gamma_k \gamma_5 \psi_h
+ {\color{black} c_{A_{k,1}}} \bar{\psi}_l \frac{1}{2} (\nabla_i^S - \overleftarrow{\nabla}_i^S) \gamma_i \gamma_5 \gamma_k \psi_h + {\color{black} c_{A_{k,2}}} \bar{\psi}_l \frac{1}{2} (\nabla_k^S - \overleftarrow{\nabla}_k^S) \gamma_5 \psi_h \nonumber \\
&\ \qquad \qquad 
+ {\color{black} c_{A_{k,3}}} \bar{\psi}_l \frac{1}{2} (\nabla_i^S + \overleftarrow{\nabla}_i^S) \gamma_i \gamma_5 \gamma_k \psi_h
+ {\color{black} c_{A_{k,4}}} \bar{\psi}_l \frac{1}{2} (\nabla_k^S + \overleftarrow{\nabla}_k^S) \gamma_5 \psi_h \big\}, \nonumber
\end{align}
and similarly for the vector current (we use the notation from Ref.\cite{leshouches}). $\psi_l$ denotes a relativistic, massless fermion, 
whereas $\psi_h$ is a nonrelativistic heavy fermion. In order to define HQET and the currents 
at the next-to-leading order one has to fix 3 parameters in $\mathcal{L}_{\HQET}$ 
and 2 $\times$ 3 parameters in $A_0(x)$ and $V_0(x)$ and 2 $\times$ 5 
in $A_k(x)$ and $V_k(x)$ giving in total 19 parameters. They are usualy denoted collectively by $\omega_i$, with $i=1, \dots, 19$. 
In this work we concentrate on the parameters $\omega_{\kin}$ and 
$c_{A_{0,1}}$ and the corresponding matching conditions.

\section{How to determine the HQET parameters?}
\label{sec2}

The HQET parameters are determined by considering observables $\phi_i$ which can be reliably calculated 
in lattice QCD and in lattice HQET. 
The matching condition reads
\begin{equation}
\phi_{i,\QCD}(L,z,a = 0) \stackrel{!}{=} \phi_{i,\HQET}(L,z,a, \{ {\color{black}\omega(z,a)} \}) = \phi_{i,\stat}(L,a) + \phi_{ij,1/m}(L,a) \ {\color{black}\omega_j(z,a)},
\label{matching}
\end{equation}
where $L$ is the size of the finite volume in which the observables $\phi_i$ are defined, $a$ is the lattice 
spacing and $z$ is a dimensionless parameter used to fix the heavy quark mass $m$ given by $z = \bar{m}(L) L$, 
where $\bar{m}(L)$ is the mass defined in the lattice minimal subtraction scheme \cite{leshouches2}. 
In Ref.\cite{3} it was proposed to use observables defined in the Schr\"odinger functional framework which 
differs from the usual one by the boundary conditions that are imposed
on the fields at time 0 and $T$ (for a more detailed discussion of the Schr\"odinger functional 
framework see \cite{leshouches2}). Apart from the usual fields in the bulk one has boundary fields which can
be used to construct correlation functions. 
The observables analyzed in this work are constructed from boundary-to-boundary or boundary-to-bulk correlation functions, e.g.
\begin{align}
F_1(\theta) &= -\frac{a^{12}}{2L^6} \sum_{\textbf{u}, \textbf{v}, \textbf{y}, \textbf{z}}\langle \bar{\zeta}'_l(\textbf{u}) \gamma_5
\zeta'_h(\textbf{v}) \bar{\zeta}_h(\textbf{y}) \gamma_5 \zeta_l(\textbf{z}) \rangle, \\
K_1(\theta) &= -\frac{a^{12}}{6L^6} \sum_i \sum_{\textbf{u}, \textbf{v}, \textbf{y}, \textbf{z}}\langle \bar{\zeta}'_l(\textbf{u}) \gamma_i
\zeta'_h(\textbf{v}) \bar{\zeta}_h(\textbf{y}) \gamma_i \zeta_l(\textbf{z}) \rangle, \\
f_{A_{0}}(\theta, x_0) &= - \frac{a^6}{2} \sum_{\textbf{u}, \textbf{v}} \langle \bar{\zeta}_h(\textbf{u}) \gamma_5 \zeta_l(\textbf{v})
\big(A_{0}\big)_I(x_0) \rangle 
\end{align}
where $\zeta$ and $\bar{\zeta}$ denote fermionic fields living on the boundary. The $\theta$ angles 
are additional kinematic parameters corresponding to the momenta of quark fields in the bulk.
The observables are defined in such a way as to cancel all renormalization factors and the angles 
can be tuned such as to minimize cut-off effects \cite{3}  
\begin{equation}
\phi_2(\theta_1, \theta_2) = \frac{1}{4} \log \frac{F_1(\theta_1)}{F_1(\theta_2)}+\frac{3}{4} \log \frac{K_1(\theta_1)}{K_1(\theta_2)}, \quad
\phi_4(\theta_1, \theta_2) = \log \frac{f_{A_0}(\theta_1,x_0=T/2)}{f_{A_0}(\theta_2,x_0=T/2)}.
\label{obs}
\end{equation}

In Ref.\cite{3} the proposed set of matching conditions was solved at tree-level
yielding the classical HQET parameters and it was checked that the $1/m_b^2$ corrections are small.
The purpose of the present study is to confirm these conclusions by a one-loop computation similar
to the one performed in \cite{dirk}. 

The one-loop contributions to the observables Eq.\eqref{obs} were calculated 
with \verb[pastor[, an automatic tool for generation and calculation of lattice Feynman diagrams \cite{pastor}. 
It is a flexible package, which takes as input the 
discretized action, the definition of the correlation function, and parameters such as $L/a$ 
and the dimensionless heavy quark mass $z$. Then, \verb[pastor[ automatically generates the Feynman 
rules corresponding to the specified action, 
all Feynman diagrams corresponding to the requested correlation function and a numerical contribution of each diagram.
The calculations were performed for the Wilson plaquette gauge action and $\mathcal{O}(a)$-improved Wilson fermions 
with two light quarks and one massive. One-loop contributions were evaluated for $\phi_{i,\QCD}$ and $\phi_{i,\stat}$. 

\section{How to estimate the quality of the observables?}
\label{sec3}

\begin{figure}
\subfigure{ \includegraphics[width=0.5\textwidth]{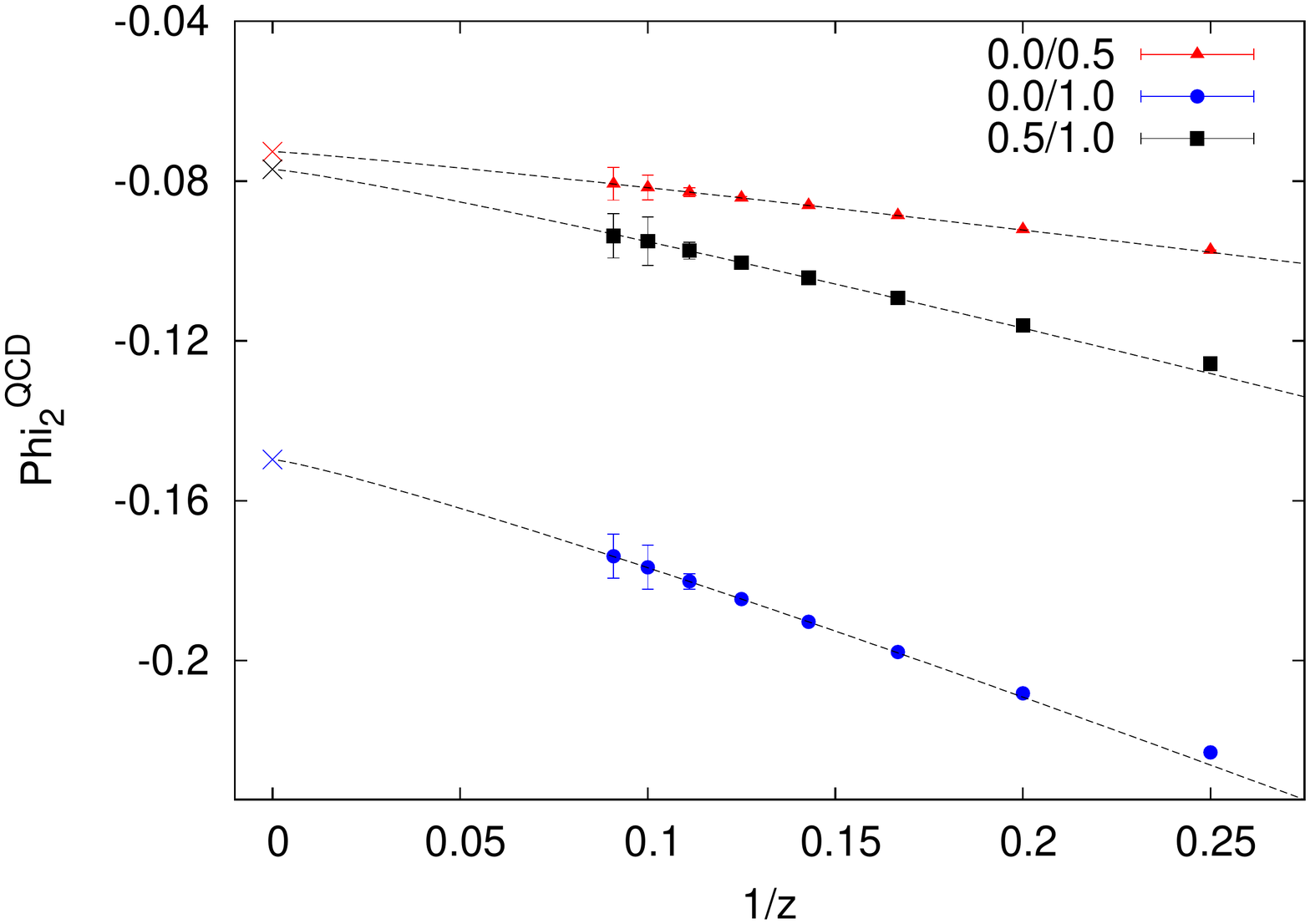} \label{r1z} }
\subfigure{ \includegraphics[width=0.5\textwidth]{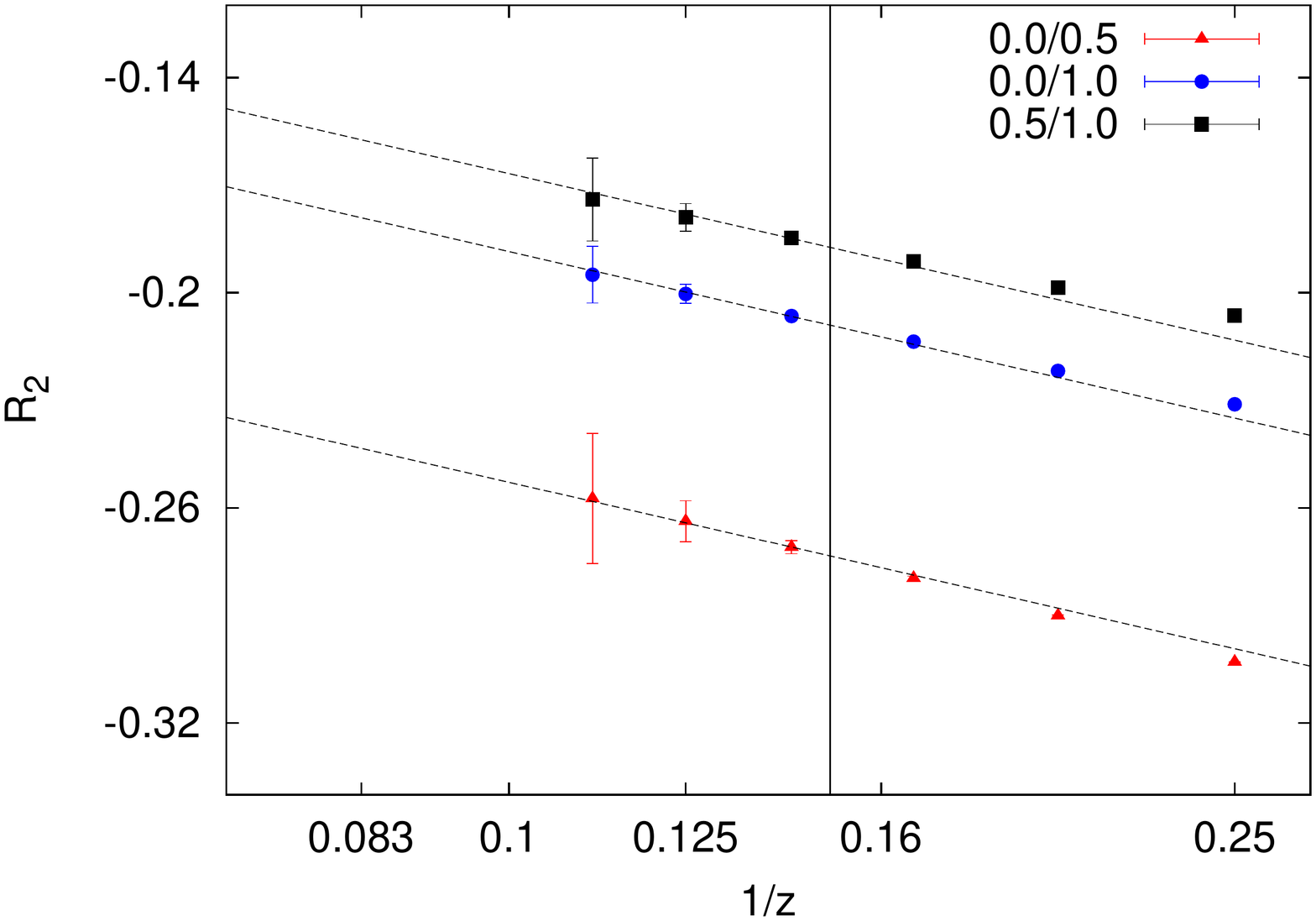} \label{r1} }
\subfigure{ \includegraphics[width=0.5\textwidth]{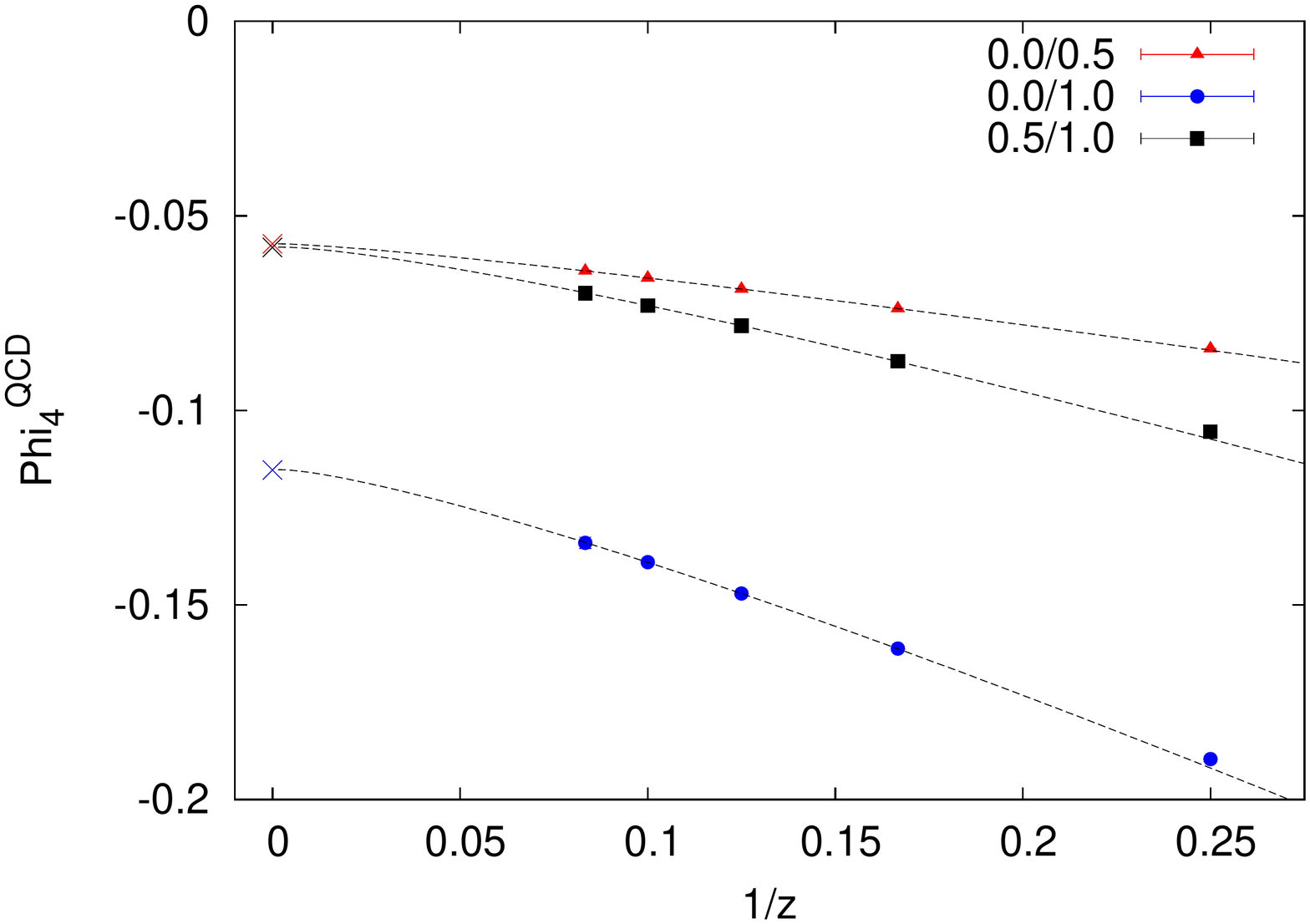} \label{r8z} }
\subfigure{ \includegraphics[width=0.5\textwidth]{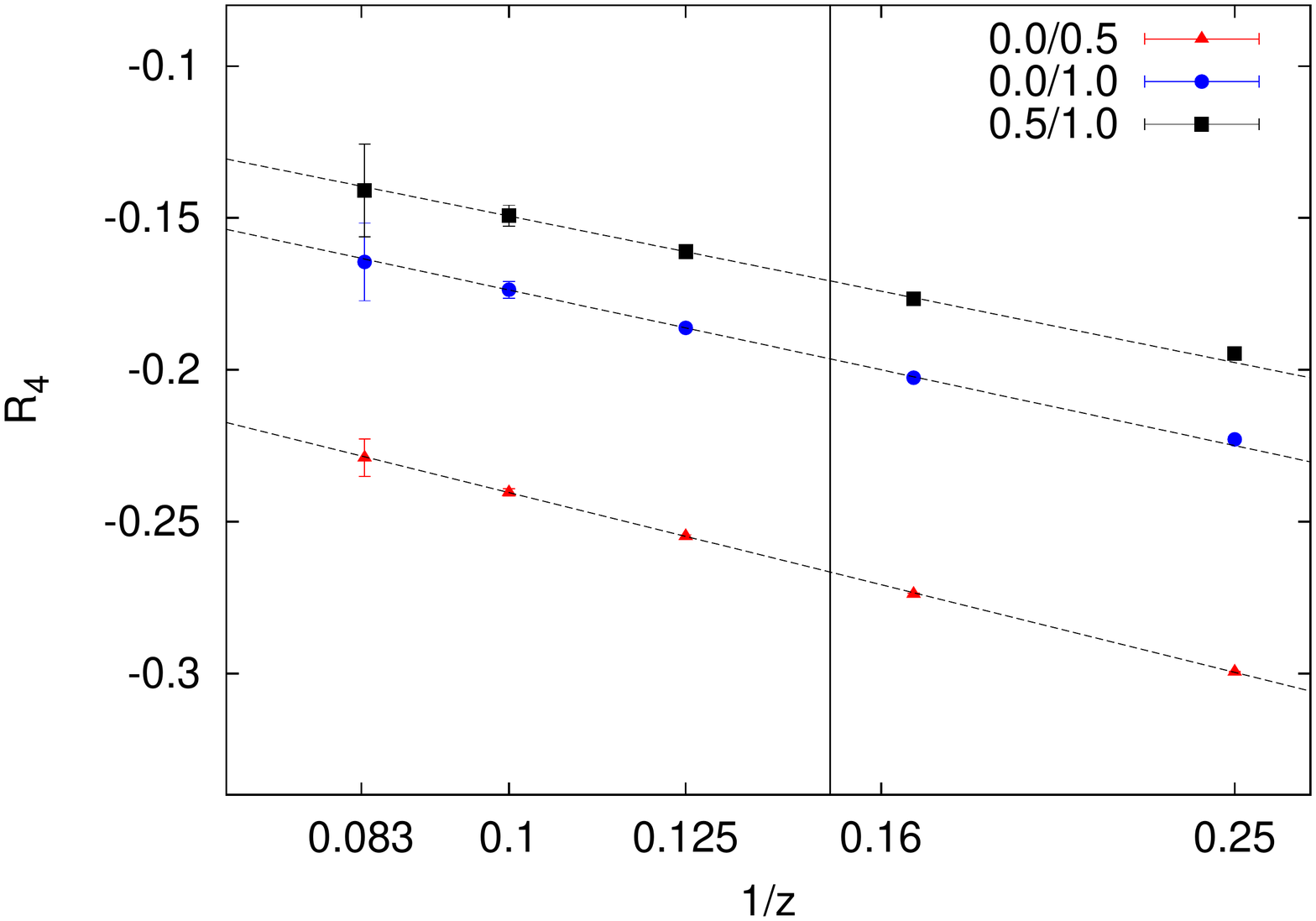} \label{r8} }
\caption{Results for $\phi_2$ (up) and $\phi_4$ (down). Figures on the left present the $z$ dependence 
of the one-loop contributions to QCD observables together with their static limit. For the fits we used the functional ansatz with 
two free parameters $f_i$ and $g_i$: $\phi^{(1)}_{i,\QCD}(z) = \phi^{(1)}_{i,\stat} + f_i/z + g_i \log(z)/z$. Figures on the 
right show the corresponding $R$ ratios. Fits were performed using data on the left of the vertical solid line.}
\end{figure}
To analyze the $1/z^{2}$ corrections to an observable $\phi$ at one-loop order in the coupling constant 
we expand the matching condition 
Eq.\eqref{matching} in $g^2$ and get (dropping terms of order $g^4$ and $1/z^{2}$)
\begin{equation}
{\color{black}\phi_{\QCD}^{(0)}(z)} + g^2 \phi_{\QCD}^{(1)}(z) = {\color{black}\phi_{\stat}^{(0)}} + g^2 \phi_{\stat}^{(1)} 
+ z^{-1} \sum_t \Big( {\color{black}\hat{\omega}_{t}^{(0)} \hat{\phi}_{t}^{(0)}} + g^2 \hat{\omega}_{t}^{(1)}(z) \hat{\phi}_{t}^{(0)} + 
g^2 \hat{\omega}_{t}^{(0)} \hat{\phi}_{t}^{(1)} \Big) 
\label{exp}
\end{equation}
where the sum over $t$ refers to different subleading contributions. 
The parameters $\hat{\omega}_t$ differ from the HQET parameters in Eq.\eqref{matching} 
by an explicit factor $1/\bar{m}(L)$ which was factored out, whereas $\hat{\phi}_t$ have an explicit factor $L$. In this notation
the kinetic contribution is $\hat{\omega}_{\kin}^{(0)} = \frac{1}{2}$, 
the spin contribution vanishes at tree level ($\phi^{(0)}_{\spin}=0$), and  
the remaining contributions correspond to corrections to the current operators proportional to the coefficients $c_X$. 
To quantify the $1/z^2$ corrections we define a ration $R$ by extracting the one-loop contribution from 
Eq.\eqref{exp} by dividing by $( {\color{black}\phi_{\QCD}^{(0)}(z) - \phi_{\stat}^{(0)}} ) g^2$
\begin{align}
R(\theta_1, \theta_2) = \frac{\phi_{\QCD}^{(1)}(z) - \phi_{\stat}^{(1)}}{{\color{black}\phi_{\QCD}^{(0)}(z) - \phi_{\stat}^{(0)}}} &= 
\frac{ \sum_t {\color{black}\hat{\omega}_{t}^{(0)}} \hat{\phi}_{t}^{(1)}(\theta_1, \theta_2)}{ \sum_t {\color{black}\hat{\omega}_{t}^{(0)}} {\color{black} \hat{\phi}_{t}^{(0)}(\theta_1, \theta_2)}} +
\frac{ \sum_t \hat{\omega}_{t}^{(1)}(z) {\color{black}\hat{\phi}_{t}^{(0)}(\theta_1, \theta_2)} }{\sum_t {\color{black}\hat{\omega}_{t}^{(0)} {\color{black}\hat{\phi}_{t}^{(0)}(\theta_1, \theta_2)}}}  \\
&= {\color{black}\alpha(\theta_1, \theta_2)  + \beta(\theta_1, \theta_2) +\gamma(\theta_1, \theta_2)\log(z)} \label{def_r}
\end{align}
Since the left hand side of Eq.\eqref{def_r} has a well defined continuum limit, the right hand side can be also considered
in the continuum (the right hand side of Eq.\eqref{def_r} must be considered as a entity since the particular terms in the 
sum may be divergent as $a \rightarrow 0$).
Note that the explicit $1/z$-dependence cancels. The only $z$ dependence remains in $\hat{\omega}_t^{(1)}(z)$ and can be parametrized as
$\hat{\omega}_t^{(1)}(z) = \beta_t + \gamma_t \log(z)$. Hence, when $R$ is plotted on a linear-log plot, the ratio $R$ measures simultaneously:
\begin{itemize}
\item \emph{$1/z^{2}$ corrections}: deviations from a linear behaviour signal $1/z^{2}$ contributions,
\item \emph{slope}: the coefficient of the subleading logarithm.
\end{itemize}
The slope gives information about a 
specific linear combination of the anomalous dimensions of the $1/m_b$ operators, which at one-loop order are 
universal (independent of the scheme) and in some cases can be predicted analytically. We will now present the ratios $R$ for two 
representative matching conditions.


The simplest matching condition is the one for $\omega_{\kin}$ \cite{3}. The $z$ dependence of the one-loop 
contribution to the observable $\phi_2$ is shown on 
figure \ref{r1z}. The ratio $R_2$ is particularly simple since there
is only one subleading contribution, namely $\hat{\omega}_{\kin}$,
\begin{align}
R_{2}(\theta_1, \theta_2) \equiv \frac{\phi_{2,\QCD}^{(1)}(z) - \phi_{2,\stat}^{(1)}}{{\color{black}\phi_{2,\QCD}^{(0)}(z) - \phi_{2,\stat}^{(0)}}} &= 
\frac{\hat{\phi}_{2,\kin}^{(1)}(\theta_1, \theta_2)}{{\color{black}\hat{\phi}_{2,\kin}^{(0)}(\theta_1, \theta_2)}} +
\frac{\hat{\omega}_{\kin}^{(1)}(z)}{{\color{black}\hat{\omega}_{\kin}^{(0)}}} 
 = {\color{black}\alpha(\theta_1, \theta_2)  + \beta + \gamma \log(z)} \label{rf1}
\end{align}
Using the fact that in the continuum $\bar{m}(L) \omega_{\kin}^{(0)} = \frac{1}{2}$  
we have $\hat{\omega}_{\kin}^{(1)}(z)/{\color{black}\hat{\omega}_{\kin}^{(0)}} = 2 \bar{m}(L)\omega_{\kin}^{(1)}(z)$. 
From continuum HQET we know (see for example \cite{grozin}) that reparametrization invariance 
fixes the renormalization factor for the kinetic 
operator to its classical value to all orders of perturbation theory. This is true if the quark mass 
used to define $\omega_{\kin}$ is given as the pole mass. In our computation we use the $\bar{m}(L)$ mass, therefore
a conversion factor needs to be included. The one-loop conversion between the pole mass and the $\overline{\textrm{MS}}$ 
scheme can be taken for example from \cite{msbar_mass} 
where its 3-loop version was derived, whereas the relation between the $\overline{\textrm{MS}}$ mass and the 
$\textrm{MS}_{\textrm{lat}}$ was given in \cite{sommer_kurth}. Hence, we
obtain the one-loop correction to $\omega_{\kin}$ as
\begin{equation}
\bar{m}(L) \omega_{\kin}^{(1)}(z) = -\frac{1}{6\pi^2} -\frac{1}{2}0.122282 \ C_F +\frac{1}{4\pi^2} \log z,
\end{equation}
from which the parameters in Eq.\eqref{rf1} can be obtained, namely, 
$\beta = -\frac{1}{3\pi^2} - 0.122282 \ C_F$ and $\gamma = \frac{1}{2\pi^2}$.
The data shown on figure \ref{r1} exhibits a slope compatible with the predicted one. We can also conclude 
that $1/z^{2}$ corrections are equally small for all sets of $\theta$ angles, hence the best setting can be 
choosen by Monte Carlo precision and tree-level considerations.



Figure \ref{r8} shows the $R$ ratio for the observable $\phi_4$, where several terms contribute 
to the sums on the right hand side of Eq.\eqref{def_r} and
the $\theta$-dependence of the coefficient of the logarithm doesn't cancel any more. Hence, a seperate fit
was performed for each set of $\theta$ angles. Again, one concludes that all data lie on straight lines and therefore
the $1/z^{2}$ correction are indeed small. 

\section{Conclusions}
Lattice HQET is a prototype of an effective theory where one can perform a non-perturbative matching. We have
studied quantitatively the contamination of the matching conditions by $1/m_b^2$ contributions and confirmed
the tree-level conclusion that such corrections are negligible. The ratios $R$ introduced in Eq.\ref{def_r} proved to be  
usefull as they provide a handle for the various parts of the $1/m_b$ contributions. Complete results for the remaining 
matching conditions will be presented elsewhere \cite{4}.

\acknowledgments
The author would like to thank especially D. Hesse for the help with \verb[pastor[ and R. Sommer and H. Simma for many useful discussions.


\begin{thebibliography}{99}
\bibitem{eichten_hill} E. Eichten, B. Hill, Phys. Lett. B 234 (1990) 511, Phys. Lett. B 243 (1990) 425,
\bibitem{grozin} A. Grozin, \emph{Heavy quark effective theory}, Springer (2004),
\bibitem{leshouches} R. Sommer, in \emph{Modern perspectives in lattice QCD}, Springer (2010),
\bibitem{sachrajda} C. Sachrajda, Nucl. Phys. Proc. Suppl. 185 (2008) 62,
\bibitem{heitger_sommer} J. Heitger, R. Sommer, JHEP 02 (2004) 022,
\bibitem{1} B. Blossier, M. Della Morte, N. Garron, R. Sommer, JHEP 06 (2010) 002, 
\bibitem{2} B. Blossier, M. Della Morte, P. Fritzsch, N. Garron, J. Heitger, H. Simma, R. Sommer, N. Tantalo, JHEP 09 (2012) 132,
\bibitem{leshouches2} P. Weisz, in \emph{Modern perspectives in lattice QCD}, Springer (2010),
\bibitem{3} M. Della Morte, S. Dooling, J. Heitger, D. Hesse, H. Simma, \emph{in preparation},
\bibitem{dirk} D. Hesse, R. Sommer, JHEP 1302 (2013) 115,
\bibitem{pastor} D. Hesse, \verb]pastor], \emph{in preparation},
\bibitem{msbar_mass} N. Gray, D. Briadhurst, W. Grafe, K. Schilcher, Z. Phys. C 48 (1990) 673,
\bibitem{sommer_kurth} M. Kurth, R. Sommer, Nucl. Phys. B 623 (2002) 271,
\bibitem{4} P. Korcyl, \emph{in preparation}.
\end{thebibliography}
\end{document}